\documentclass[10pt,journal,compsoc]{IEEEtran}

\usepackage{amsmath,amsfonts}
\usepackage{algorithmic}
\usepackage{algorithm}
\usepackage{array}
\usepackage[caption=false,font=normalsize,labelfont=sf,textfont=sf]{subfig}
\usepackage{color, soul}
\usepackage{textcomp}
\usepackage{stfloats}
\usepackage{url}
\usepackage{verbatim}
\usepackage{graphicx}
\usepackage{cite}
\usepackage{ragged2e}
\usepackage{hyperref}
\newcommand{\para}[1]{\vspace{.05in}\noindent\textbf{#1}}

\begin{document}

\title{Multi-sensor Learning Enables Information Transfer across Different Sensory Data and Augments Multi-modality Imaging}

\author{Lingting Zhu, Yizheng Chen, Lianli Liu, Lei Xing, and Lequan Yu
\thanks{L. Zhu is with the Department of Statistics and Actuarial Science, The University of Hong Kong, Hong Kong SAR, China.}

\thanks{Y. Chen, L. Liu, and L. Xing are with Department of Radiation Oncology, Stanford University, 94305, CA, USA. (e-mail: lei@stanford.edu).}
\thanks{L. Yu is with the Department of Statistics and Actuarial Science, The University of Hong Kong, Hong Kong SAR, China and also with the Department of Radiation Oncology, Stanford University, 94305, CA, USA. (e-mail: lqyu@hku.hk). Corresponding author: Lei Xing, Lequan Yu.}
}

\markboth{Journal of \LaTeX\ Class Files,~Vol.~14, No.~8, August~2021}%
{Shell \MakeLowercase{\textit{et al.}}: A Sample Article Using IEEEtran.cls for IEEE Journals}


\IEEEtitleabstractindextext{
\begin{abstract}
    \justifying
    Multi-modality imaging is widely used in clinical practice and biomedical research to gain a comprehensive understanding of an imaging subject. Currently, multi-modality imaging is accomplished by post hoc fusion of independently reconstructed images under the guidance of mutual information or spatially registered hardware, which limits the accuracy and utility of multi-modality imaging. Here, we investigate a data-driven multi-modality imaging (DMI) strategy for synergetic imaging of CT and MRI. We reveal two distinct types of features in multi-modality imaging, namely intra- and inter-modality features, and present a multi-sensor learning (MSL) framework to utilize the crossover inter-modality features for augmented multi-modality imaging. The MSL imaging approach breaks down the boundaries of traditional imaging modalities and allows for optimal hybridization of CT and MRI, which maximizes the use of sensory data. We showcase the effectiveness of our DMI strategy through synergetic CT-MRI brain imaging. The principle of DMI is quite general and holds enormous potential for various DMI applications across disciplines.
    \end{abstract}
    
} 
\maketitle


\section{Introduction}
\IEEEPARstart{T}{he} success of clinical decision-making and scientific inquiry heavily relies on comprehensive imaging techniques for visualizing inner body characteristics. Many imaging modalities, such as CT and MRI, have been developed over the years to probe the imaging subject. In reality, no single modality can provide complete information to meet all clinical demands – multi-modality imaging has become sought after to combine the complementary information of two or more imaging modalities and push the frontier of biomedical imaging to gain a holistic view. Unfortunately, multi-modality imaging today~\cite{multi-1,multi-2,multi-3} is achieved by post hoc fusion of independently acquired images, guided by mutual information~\cite{b1,b2,b3,b4}. Typically, existing researches concentrates on the fusion of well-reconstructed modalities.

Given the unique capability of deep learning to extract and integrate knowledge from diverse sources~\cite{b3,b4,b5,b6,b7,b8,b9,b10,b11,b12,b13,b14,b15,b16,b17}, a fundamental question we ask here is whether the data acquired by sensors of two (or more) modalities can be processed synergistically so that (i) the image reconstruction of one modality can be enhanced or even completely done by the sensory data from another modality, and (ii) the resultant pixelated image information can be locally or globally tuned and optimized to provide on-demand hybridization of different modalities by varying the proportionality of the sensory data of different modalities. In this way, the boundaries of conventional imaging modalities can be removed to lend information that would otherwise be impossible to attain.

\begin{figure*}[t]
\centering
\includegraphics[width=0.95\textwidth]{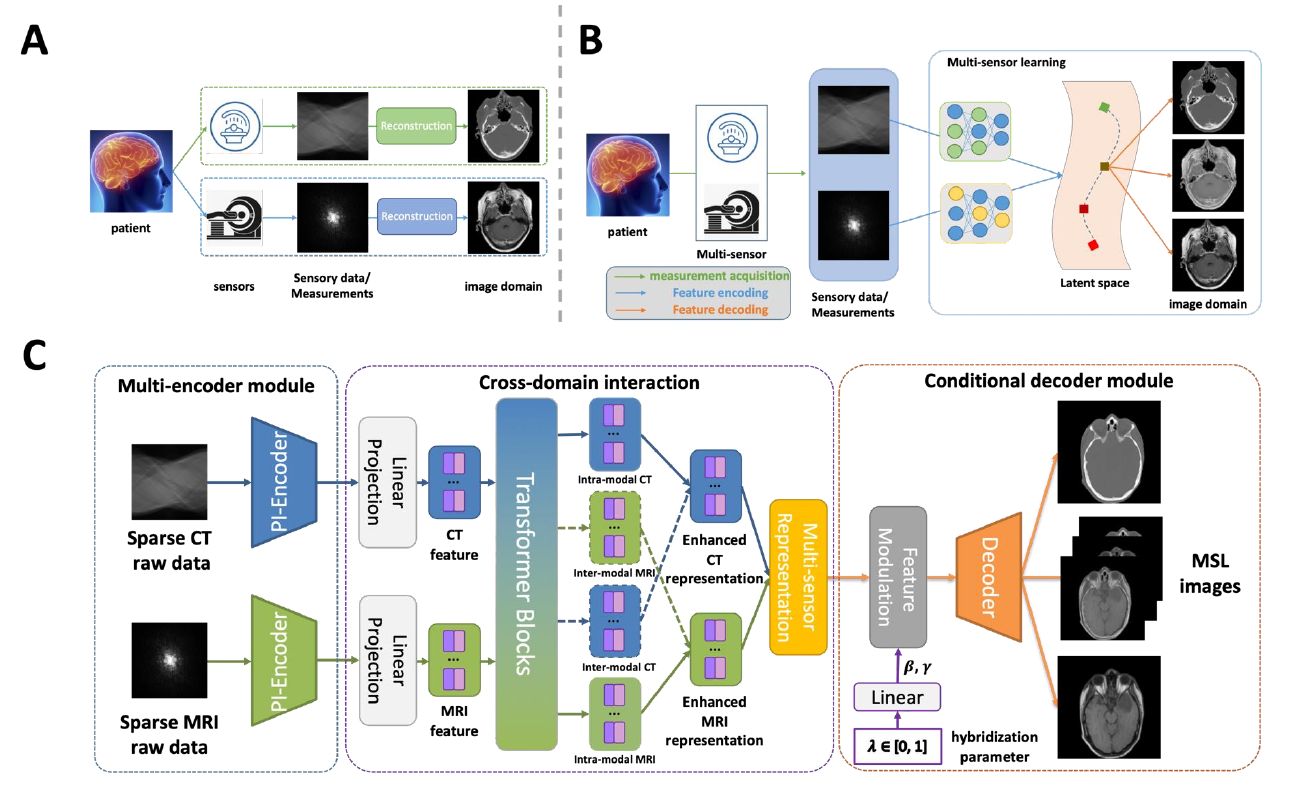}
\caption{\textbf{Data-driven multi-modality imaging with multi-sensor learning.} 
\textbf{(A)} Conventional approach reconstructs independent images of different modalities and fuses them post hoc without any information transfer across different modalities. \textbf{(B)} The proposed DMI strategy enables synergetic imaging by sharing inter-modality features in a latent representation space. \textbf{(C)} Schematic diagram of our MSL framework, featuring a multi-domain encoder-decoder design and a transformer-based cross-domain interaction module that extracts and transfers crossover inter-modality information between the two modalities. The cross-domain interaction module obtains intra- and inter-modality CT and MRI representations to provide a multi-sensor representation. A sequence of MSL images is generated by transforming the multi-sensor representation with a conditional decoder module.} 
\label{overview}
\end{figure*}

In this work, we investigate a data-driven multi-modality imaging (DMI) strategy and propose a broadly applicable multi-sensor learning (MSL) framework for the synergetic imaging of CT and MR, see Fig.~\ref{overview}A \& B. 
Particularly, we design a multi-domain encoder-decoder architecture~\cite{b3,b19} (Fig.~\ref{overview}C), where physics-informed (PI) encoders (i.e., an encoder with the incorporation of the analytical back-transform of the physical forward function of the imaging modality) are implemented to extract sensory features from the respective sensory data. A cross-domain interaction module is then introduced to couple sensory features into multi-sensor representation in the latent space by using a transformer network~\cite{b20}, followed by a decoder to transform the multi-sensor representation into a series of MSL images with different levels of mixing between the two modalities. Our work reveals that there are generally two distinct types of features in DMI, i.e., intra- and inter-modality features, and highlights the potential of the crossover inter-modality features extracted by MSL to enable the information transfer across modalities so that we can enhance or even generate images from the sensory data of alternative modality (Fig.~\ref{overview}B). Existing works~\cite{jointspace-1, jointspace-2, jointspace-3} in multi-modality primarily focus on joint learning across two or more modalities, typically including visual and text information, and assume that information from different modalities can be mapped onto shared or correlated subspaces. In our work, we model multi-sensor learning by jointly learning a latent space from raw sensory data, thereby enabling multi-modality imaging from a unified representational space. To the best of our knowledge, this is the first investigation of a learning-based integration of multi-sensory data for synergetic imaging.

A key component of our MSL approach is the cross-domain interaction module (Fig.~\ref{overview}C), which employs transformer blocks to separate the sensory features of CT and MRI into intra- and inter-modality features. After the MSL model is trained, the sensory representation of each modality is augmented by the inter-modality feature information from the other modality. Via the cross-domain interaction module, we obtain a multi-sensor representation that enhances the CT and MRI sensor representations. It is intriguing that our MSL model can produce both CT and MRI images from even a single modality input. Moreover, our framework is capable of providing a sequence of MSL images that correspond to different mixtures of CT and MRI components by adjusting a hybridization parameter $\lambda$, which controls the proportionality of multiple sensory data. We achieve this by using the conditional instance normalization (CIN)~\cite{b21} technique, which is implemented in the conditional decoder module (Fig.~\ref{overview}C). Importantly, as $\lambda$ can be varied spatially, the tuning of modality proportionality can be done globally or locally to provide on-demand hybridization images of CT/MRI. Furthermore, an optimal MSL image with a maximal amount of information can be obtained by optimizing the entropy of the resultant image with respect to the spatial distribution of $\lambda$.

We demonstrate the potential of the proposed DMI approach with synergetic CT-MRI brain imaging using sparse CT and MRI sensory data. The concept and implementation here are quite general and can be extended to other multi-modality imaging or multi-sensor applications, such as multi-omics analysis, multi-sensory information integration in remote sensing, autonomous driving, robotics, astronomy, and so forth. As an illustration of the broad applicability of the approach, we also show the feasibility and effectiveness of MSL imaging on synergetic $T_1$- and $T_2$-weighted MRI. 
The source code is available at \url{https://github.com/HKU-MedAI/MSL}.

\para{Related Works.} 
In the past decades, a significant line of research in DMI is focused on post hoc fusion of multi-modality medical images after image reconstruction. Deep learning models have been employed for image fusion~\cite{ram2017deepfuse, liu2017multi, li2018infrared}, medical image fusion and reconstruction~\cite{multi-1, multi-2, fusion-transformer-1, fusion-transformer-2, fusion-1, fusion-2, xing2024accelerated, li2024dual, xing2024progressive}. Various network architectures have been investigated for the purpose. Noteworthy, works such as~\cite{multi-3, fusion-3, fusion-4} examined the extraction of multi-modal features through collaborative learning or by modeling cross-modality shallow features, providing comprehensive insights into the inherent relationships between different imaging modalities. Beyond spatially alignment of medical images, many efforts are devoted in aligning or fusing information from multiple medical modalities, such as images, radiology reports, and genetic data~\cite{shin2016interleaved, zhou2022generalized, wang2022multi, taleb2022contig}. In recent years, multi-modality learning has garnered significant attention in artificial intelligence~\cite{jointspace-1, jointspace-2, jointspace-3}, with efforts to model joint spaces through feature space alignment aided by large-scale multi-modality datasets. The investigation of multi-modality feature extraction and fusion has been widely integrated into various computer vision tasks~\cite{zhang2019robust, wei2020multi, xu2020discriminative, man2023bev, fusion-4}.

\section{Method}

\subsection{Overview and Problem Formulation}
Training a deep learning model requires a large amount of labeled data, which is often the bottleneck in establishing a robust model~\cite{b17,b41}. Instead of measuring a large amount of paired multi-modality sensory data and ground truth images, we digitally produce CT and MRI datasets from paired CT and MRI volumes by using a geometry consistent with our clinical scanner. Given the under-sampled raw CT sensory data $Y_{CT}$ and MRI sensory data $Y_{MRI}$, we aim to reconstruct a series of MSL images $X_{MSL}$, including the single-modality MSL-CT $X_{CT}$ and MSL-MRI $X_{MRI}$. To enable this, we assume that each MSL image (including the MSL-CT and MRI images) can be reconstructed from a latent code in the shared feature space. We, thus, propose a novel MSL framework, including a multi-domain encoder module, cross-domain interaction module, and conditional decoder module (Fig.~\ref{overview}C), to learn the latent code and reconstruct images.

\subsection{Multi-domain Physics-informed Encoder}
We employ two different physics-informed encoders (i.e., PI-Encoders) $\rm{E}_{CT}$ and $\rm{E}_{MRI}$ to extract the features of the raw CT and MRI sensory data. To ease the learning and accelerate the training of the encoders, we also incorporate the well-known FBP operator and the inverse Fourier transform operator into the encoders. The acquired sensor features $f_{CT}$ and $f_{MRI}$ can be represented as 
\begin{align} 
f_{CT}= \rm{E}_{CT} (X_{CT'}) = \rm{E}_{CT} ({\rm FBP} (Y_{CT})),\\
f_{MRI}=\rm{E}_{MRI}(X_{MRI'}) = \rm{E}_{MRI} ({\rm Fourier}^{-1} (Y_{MRI})).
\end{align}

\subsection{Cross-domain Interaction}
We obtain the coupled multi-sensor representation by forwarding the CT sensory feature $f_{CT}$ and MRI sensory feature $f_{MRI}$ into a cross-domain interaction module. The proposed cross-domain interaction module is composed of several transformer blocks and feature fusion operators. With a specially designed transformer-based token learning mechanism, the relationship and crossover inter-modality information of different sensory features are learned and thus a coupled multi-sensor representation is acquired for subsequent processing. 

\subsubsection{Sensor feature tokenization}
As each raw sensory data of CT or MRI is under-sampled and only contains partial information about the object, it is important to effectively combine the complementary information to learn a good latent code in the shared feature space. Since the paired sensory data from CT and MRI pertain to the same patient, the shared space is constructed to encode shared information as well as the complementary information acquired via distinct physical projections. Motivated by the success of transformer architecture in modeling and learning the relationship of different components in natural language processing and computer vision, we propose to employ a transformer-based module to formulate and learn the relationship of different sensory representations, and thus extract and transfer the crossover information. Specifically, after the feature maps of each modality $f_{CT}$ and $f_{MRI}$ are obtained, they are individually tokenized. We treat the feature maps as images, crop the feature maps into multiple feature patches (aka, tokens) along the spatial dimension, and adopt linear projections to get the feature tokens like ViT~\cite{b20}.

\subsubsection{Transformer-based token learning guides crossover information transfer}
To learn and transfer the crossover inter-modality information among different modalities, we design a token learning mechanism based on the Transformer blocks. After obtaining the sensory feature tokens, we first duplicate the feature tokens of each modality into two groups and then concatenate the four groups of feature tokens from two modalities as the input of the transformer blocks. The class token and the position embedding in the original ViT are kept, and they serve as the trainable parameters in our setting. Attention blocks and multiple linear projection (MLP) layers are used in the transformer blocks. In Fig.~\ref{overview}C, we show that, taking four groups of feature tokens as the input, the transformer blocks output four groups of features, $f_{intra\_CT}$, $f_{inter\_CT2MRI}$,  $f_{inter\_MRI2CT}$, and  $f_{intra\_MRI}$. The intra-modality features are trained to gain more fine-grained information about the corresponding modality, while the inter-modality features are trained to extract the overlapped information from the other modality.  We further stack several self-attention modules and MLPs in Transformer blocks to facilitate module learning.

\subsubsection{MSL imaging with sensory data of a single modality}
With well-trained intra-modality and inter-modality features, our framework can generate both MSL-CT and MSL-MRI images, as well as other MSL images with a tunable mixture of CT and MRI.  Generally, to obtain MSL images, we first obtain the enhanced CT and MRI representations $f_{enhanced\_CT}$ and $f_{enhanced\_MRI}$, by compositing the intra-modality and inter-modality features:
\begin{align} 
f_{enhanced\_CT}=f_{intra\_CT}+f_{inter\_MRI2CT}, \\
f_{enhanced\_MRI}=f_{intra\_MRI}+f_{inter\_CT2MRI}.
\end{align} 
For multi-modality cases, the $f_{enhanced\_CT}$ and $f_{enhanced\_MRI}$ are served as input to the conditional decoder module. For cases with only a single-modality input, we have only the sensory feature of one modality. For example, when there is only MRI sensory data available as input, the transformer-based token learning process would only produce $f_{intra\_MRI}$ and $f_{inter\_MRI2CT}$. In this case, the inter-modality feature fills the void of missing modality representation, we have
\begin{align} 
f_{enhanced\_CT}=f_{inter\_MRI2CT}, \\
f_{enhanced\_MRI}=f_{intra\_MRI}.
\end{align} 
In our MSL, the coupled multi-sensor representation $f_{multi\_sensor}$ are obtained by fusing the enhanced representations and an upsampling operator,
\begin{align} 
f_{multi\_sensor}={\rm UP}(f_{enhanced\_CT}+f_{enhanced\_MRI}),
\end{align} 
where $\rm UP(\cdot)$ denotes upsampling operation with transposed convolutional blocks to map the feature maps into the same dimension with the sensor features. With this cross-domain interaction scheme, our framework can learn and extract the crossover inter-modality information among different modalities and enable MSL imaging from both multi-modality and single-modality inputs.

\subsection{Conditional Decoder for MSL Image Reconstruction}
We design a shared conditional decoder module for MSL imaging. At the inference phase, we feed the coupled multi-sensor presentation and an adjustable parameter $\lambda$ (i.e., hybridization parameter) into the shared decoder to reconstruct the MSL images. The $\lambda$ is used to tune the level of hybridization. When $\lambda$ takes a value of 0 or 1, an MSL-CT or MSL-MRI image is generated. By tuning the value of $\lambda$ between 0 and 1, the decoder can output MSL images with different levels of hybridization of CT and MRI. This conditional decoding process can be formulated as,
\begin{align} 
X_{MSL, \lambda}={\rm D}(f_{multi\_sensor}, \lambda).
\end{align} 
Particularly, we can get the reconstructed MSL-CT and MSL-MRI images by adjusting the corresponding $\lambda$,
\begin{align} 
X_{CT} = X_{MSL, \lambda=0}={\rm D}(f_{multi\_sensor}, \lambda=0),\\
X_{MRI} = X_{MSL, \lambda=1}={\rm D}(f_{multi\_sensor}, \lambda=1).
\end{align} 
Specifically, for a given $\lambda$, we use a linear layer $f_s$ to get a conditional embedding $s=f_s (\lambda)$ with size of 256. We then use another two different linear projection layers to obtain the learnable 1D affine parameters  $\gamma(s)$, $\beta(s)$, which have the same size as the channel dimension of feature map $x$.
After that, these affine parameters are used to modulate the feature map $x$ in a channel-wise manner, following the CIN technique
\begin{align} 
{\rm CIN}(x;s)=\gamma(s)\left( \frac{x-\mu(x)}{\sigma (x)}\right)+\beta (s),
\end{align} 
where $\mu(x)$ and $\sigma(x)$ denote the mean and standard deviation of the feature map $x$.

Besides tuning the level of hybridization of the entire image, our approach also allows local tuning of imaging contrast. This is achieved by letting $\lambda$ to vary spatially. In detail, the $\lambda$ is allowed to a spatial distribution map $\hat{\lambda} =\{\hat{\lambda}_{i,j}\}$ with the same spatial size as the feature map. For a 2D $\hat{\lambda}$ map, we compute the corresponding 2D spatial conditional embedding $\hat{s} =\{\hat{s}_{i,j}\}$, where $\hat{s}_{i,j}=f_s(\hat{\lambda}_{i,j})$ and $\hat{\lambda}_{i,j}$ is treated as a regular hybridization parameter. Like the above globally tunable case, we obtain the two sets of spatial affine parameters $\gamma(\hat{s}) =\{\gamma(\hat{s}_{i,j})\}$, and $\beta(\hat{s}) =\{\beta(\hat{s}_{i,j})\}$ to spatially modulate the feature maps. With the same CIN technique and the expanded 2D spatial affine parameters, MSL is enabled to reconstruct images with locally tunable imaging content, without re-training or re-designing the training procedure.

To automatically obtain the optimized spatial distribution map $\hat{\lambda} =\{\hat{\lambda}_{i,j}\}$, we formulate the following optimization task: 
\begin{align} 
\max_{\hat{\lambda}} \frac{1}{2} {\rm MI}(\hat{X}_{MSL, \hat{\lambda}}, X_{CT})&+\frac{1}{2} {\rm MI}(\hat{X}_{MSL, \hat{\lambda}}, X_{MRI})-\alpha {\rm TV}(\hat{\lambda}), \\
\hat{X}_{MSL, \hat{\lambda}}&={\rm D}(f_{multi\_sensor}, \hat{\lambda}).
\end{align}
$\hat{X}_{MSL, \hat{\lambda}}$ is the MSL image with locally tunable imaging contrast for a spatial distribution map $\hat{\lambda}$, MI represents the mutual information of two given images, and TV denotes the total variation~\cite{b42} as a regularization to optimize a spatially smooth map. We use $\alpha$ as a hyperparameter to control the regularization and set it as 0.001 in our implementation. In other words, we aim to find an optimal spatial distribution map $\hat{\lambda}$ that can simultaneously maximize the MI between the resultant MSL image with MSL-CT and MSL-MRI given the trained model. Computationally, we first initialize a spatial distribution map $\hat{\lambda}^{(0)}$ and employ the gradient descent (GD) algorithm to solve the above optimization task.

\subsection{Objective Function and Training Procedure}
When training the network, we feed a minibatch of CT data and a minibatch of MRI data into the MSL model and calculate the reconstruction loss between reconstructed images and underlying ground truth images. The reconstruction loss composed of MAE items is defined as,
\begin{align} 
L_{rec}= {\rm MAE}(X_{CT},X_{CT}^{gt}) + {\rm MAE}(X_{MRI},X_{MRI}^{gt}),
\end{align}
where $X_{CT}^{gt}$ and $X_{MRI}^{gt}$ represent the ground truth CT and MRI images, respectively. Additionally, we add a fusion loss to further smooth the reconstructed images in the sequence. We select the central MSL image reconstructed by the decoder as an anchor image when the adjustable parameter $\lambda$ is set as 0.5. Mean squared error (MSE) is utilized to compute the fusion loss
\begin{align} 
L_{fusion}=&\frac{1}{2} {\rm MSE}(X_{CT},X_{MSL,\lambda=0.5})\nonumber\\+&\frac{1}{2}{\rm MSE}(X_{MRI},X_{MSL,\lambda=0.5}).
\end{align}
The fusion loss is defined as the average MSE towards CT and MRI ground truth to guide the anchor image to locate at a proper level in the series of MSL images. As the ground truth for a hybrid image does not exist, we leverage the single-modality CT and MRI images to guide the training of the MSL model and the effectiveness of this fusion loss strategy is demonstrated by our experimental results.

To encourage the intra-modality feature to gain more fine-grained information about the corresponding modality and the inter-modality feature to extract the overlapped information from the other modality, we also design an auxiliary loss scheme to guide the framework training. As mentioned earlier, the enhanced CT and MRI representations will degrade when only single-modality sensory data is available. Besides obtaining the regular multi-sensor representation $f_{multi\_sensor}$, we also simulate a single-modality sensory data scenario and obtain an auxiliary multi-sensor representation $f_{multi\_sensor}^{aux}$ with only one intra-modality feature and one inter-modality feature. We then generate the respective MSL-CT and MRI images $X_{CT}^{aux}$ and $X_{MRI}^{aux}$ from the auxiliary multi-sensor representation with the decoder module
\begin{align} 
X_{CT}^{aux}=X_{MSL, \lambda=0}^{aux}={\rm D}(f_{multi\_sensor}^{aux},\lambda=0),\\
X_{MRI}^{aux}=X_{MSL, \lambda=1}^{aux}={\rm D}(f_{multi\_sensor}^{aux},\lambda=1).
\end{align}

After that, we adopt an auxiliary reconstruction loss $L_{aux}$ to enable imaging from the single-modality sensory data, and an auxiliary feature loss $L_{aux\_feat}$ to encourage the inter-modality feature to extract the overlapped information from the other modality 
\begin{align} 
L_{aux}= &{\rm MAE}(X_{CT}^{aux},X_{CT}^{gt}) \nonumber\\+ &{\rm MAE}(X_{MRI}^{aux},X_{MRI}^{gt}),
\end{align}
\begin{align} 
L_{aux\_feat} = &{\rm MAE}(f_{intra\_CT},f_{inter\_MRI2CT}) \nonumber\\+ &{\rm MAE}(f_{intra\_MRI},f_{inter\_CT2MRI}).
\end{align}

Finally, we use hyperparameters $\phi_f$, $\phi_a$, and $\phi_e$ to balance the weights of the above loss items, and the total objective function to train the MSL framework is defined as
\begin{align} 
L= L_{rec}  +\phi_f L_{fusion}+\phi_a L_{aux}+\phi_e L_{aux\_feat}.
\end{align}

\begin{figure*}[t]
\centering
\includegraphics[width=0.7\textwidth, height=0.95\textwidth]{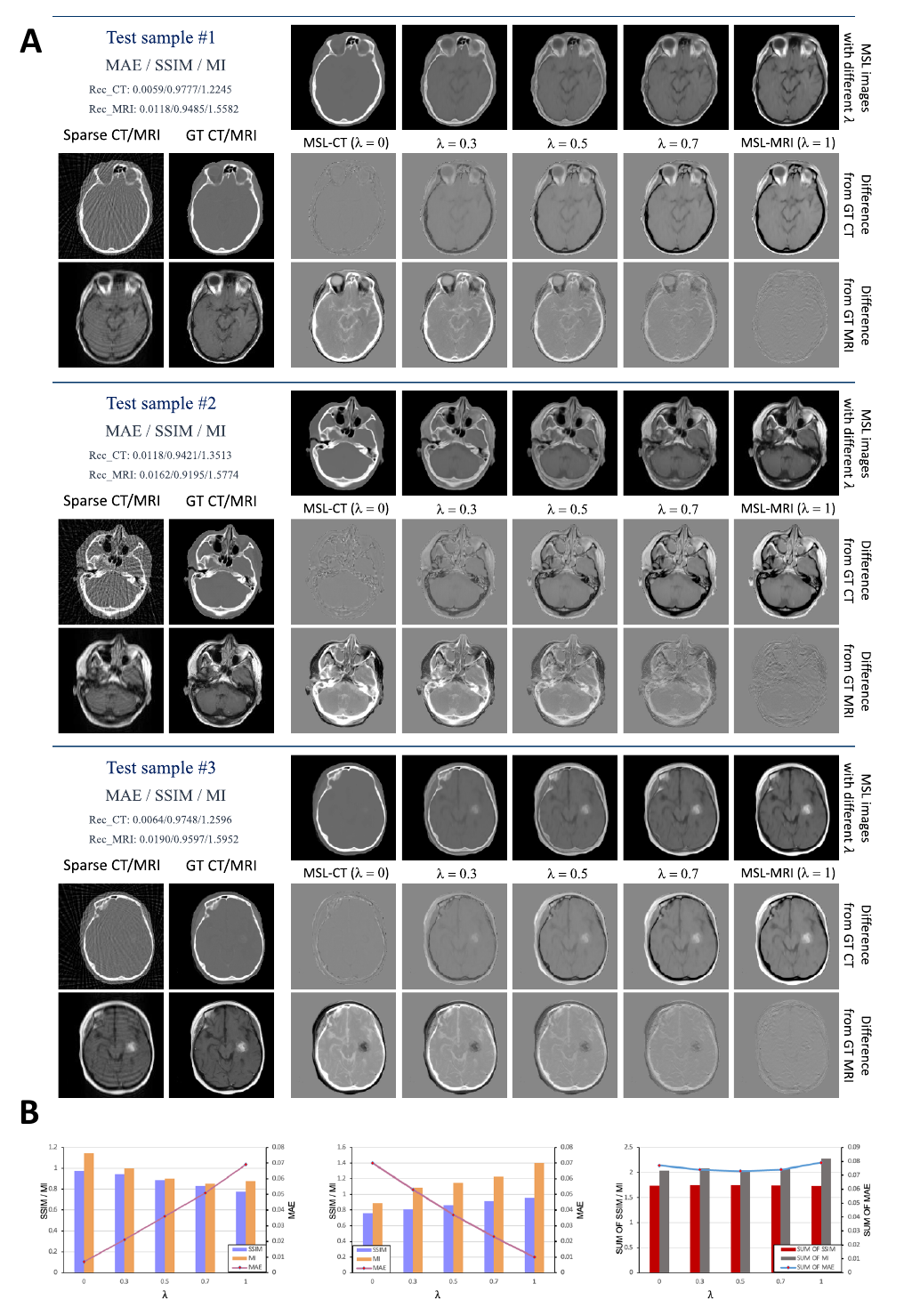}
\caption{\textbf{Qualitative and quantitative analyses of the MSL images.} 
\textbf{(A)} MSL imaging results of three different samples. Sparse images from conventional reconstruction algorithms, ground truth (GT) images, MSL images, and difference images between the MSL image and GT CT and MRI images are shown. \textbf{(B)} Quantitative evaluation of MSL imaging. The MAE, SSIM, and MI metrics of the MSL image with respect to the GT CT and MRI images at different $\lambda$ are shown in the left and middle panels, respectively. The right panel shows the summation of the paired metrics, which remains almost unchanged as $\lambda$ changes.} 
\label{results}
\end{figure*}

\begin{figure*}[t]
\centering
\includegraphics[width=0.75\textwidth, height=0.5\textwidth]{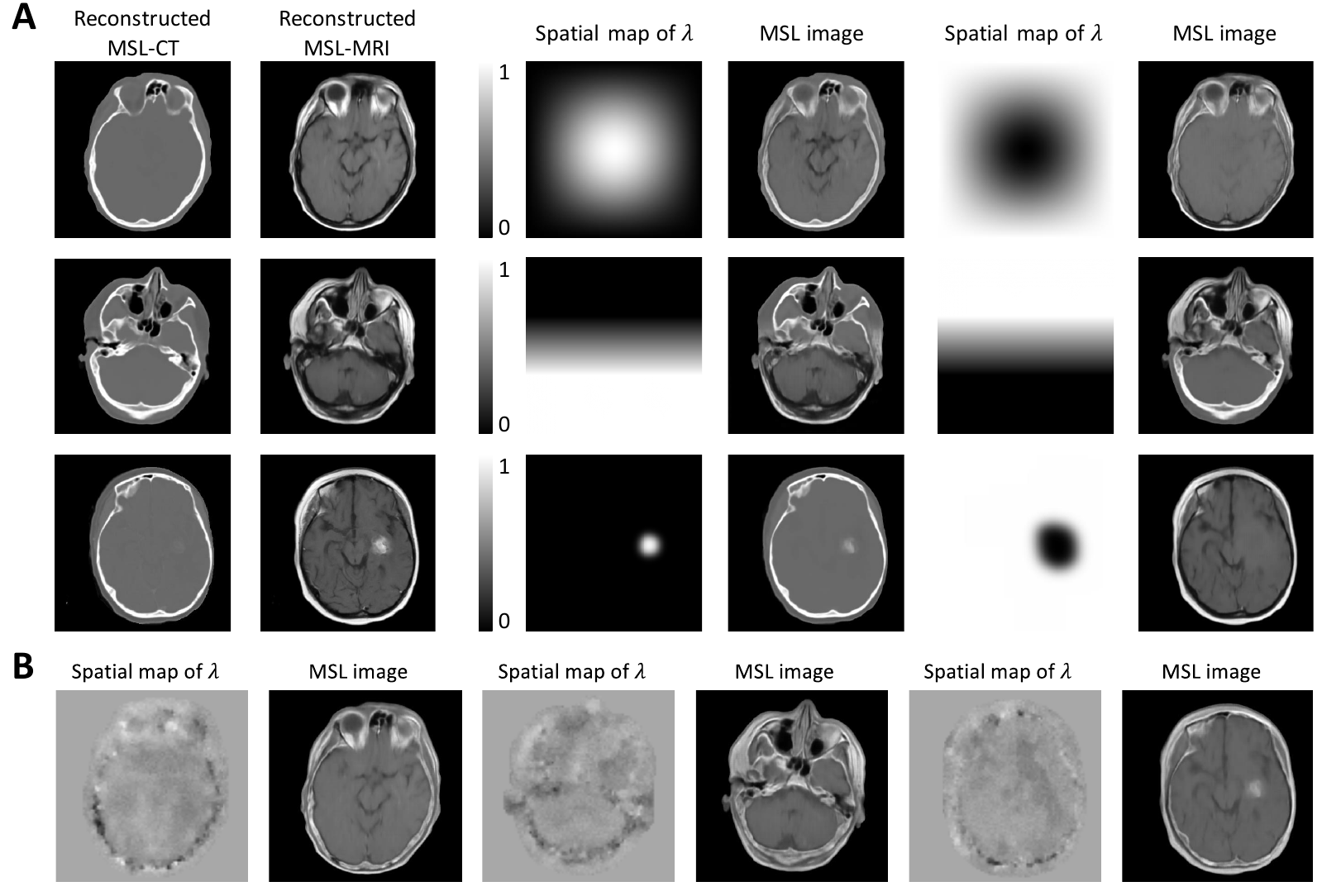}
\caption{\textbf{MSL imaging with local contrast tuning.} 
\textbf{(A)} The left panel displays the MSL-CT ($\lambda=0$) and MSL-MRI ($\lambda=1.0$) images for three different samples. The corresponding MSL images generated with six different spatial $\lambda$ distributions are shown in the middle and left panels (where the white and black denote $\lambda$=1 and 0, respectively). \textbf{(B)} The optimized spatial maps of $\lambda$ and the corresponding MSL images for the three samples shown in A are presented.} 
\label{lambda}
\end{figure*}

\subsection{Network Architecture Details}
The PI encoder consists of the respective back-transform operator (e.g., FBP or inverse Fourier transform) to incorporate the measurement-image geometry~\cite{b43}, three stridden convolutional blocks to reduce the spatial dimension of the feature maps, and three residual blocks to process the extracted features further. Similarly, the decoder part consists of three residual blocks, two upsampling blocks, and one convolutional layer to reconstruct the MSL images from the coupled multi-sensor representation. In our implementation, we use instance normalization layers and ReLU activations. In the decoder, the proposed CIN technique is used in the residual blocks and upsampling blocks. Specifically, c7s1-$k$ represents a 7×7 convolutional block with $k$ filters and stride 1; d$k$ denotes a 4×4 convolutional block with $k$ filters and stride 2; R$k$ denotes a residual block that contains two 3×3 convolutional blocks with $k$ filters and stride 1; u$k$ denotes a 3×3 transposed convolutional block with $k$ filters and stride 2. The detailed architecture of the encoder and decoder can be written as,
\begin{enumerate}
    \item Encoder: c7s1-64, d128, d256.
    \item Decoder: R256, u128, u64, c7s1-1.
\end{enumerate}
For the transformer-based cross-domain interaction module, we crop the feature maps as 4×4 patches in the spatial dimension and use a linear projection layer to project the feature tokens to the dimension of the Transformer blocks, which is set to 1,024. We then use two Transformer blocks, each composed of a self-attention block and an MLP. The module finally outputs $f_{multi\_sensor}$ by upsampling the fused enhanced features with u512 and u256 blocks.

\begin{figure*}[t]
\centering
\includegraphics[width=0.65\textwidth, height=0.55\textwidth]{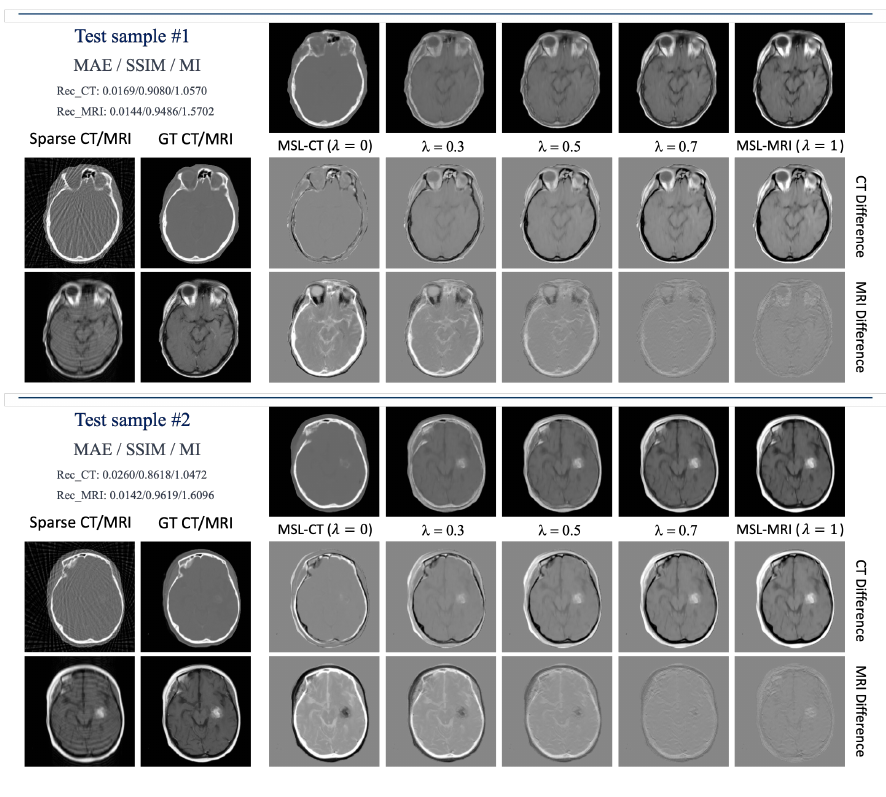}
\caption{\textbf{MSL imaging results with only MRI sensory data input.} 
Even with only MRI k-space data as input during testing, MSL is able to generate a continuum of images with adjustable hybridization of MRI and CT imaging contents.} 
\label{onlymri}
\end{figure*}

\begin{figure*}[t]
\centering
\includegraphics[width=0.65\textwidth, height=0.28\textwidth]{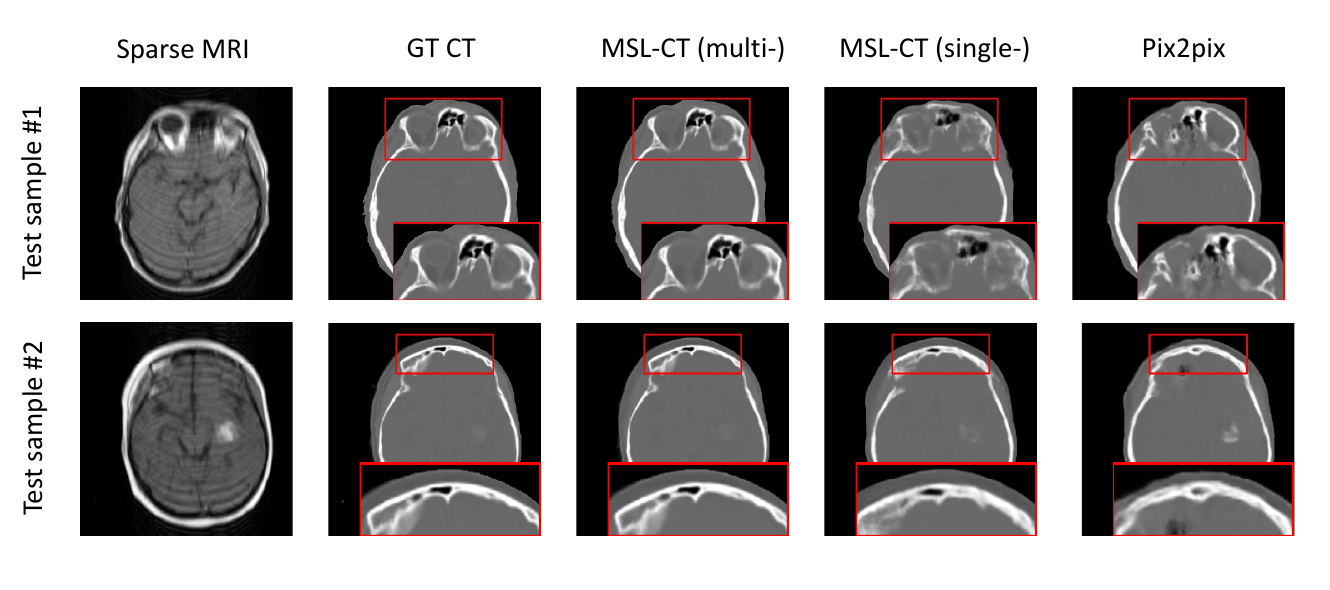}
\caption{\textbf{MSL imaging results compared with the Pix2pix image translation method.} 
MSL imaging results compared with the Pix2pix image translation method. The outcomes of MSL from both multi-modality and single-modality inputs and that of Pix2pix are present in different columns.}
\label{translation}
\end{figure*}

\section{Experiments}

\subsection{Datasets and Implementation Details}

The training data include pairs of CT and MRI images from the RIRE\footnote{\url{https://rire.insight-journal.org/}} project~\cite{b44}, which are acquired for the same anatomy and are both sized to 384×384. The under-sampled CT sinogram is produced via a forward projection from 64 different directions out of a full rotation ranging from 0 to 360 degrees, and the under-sampled MRI $k$-space data is generated by applying the Fourier transform to the original image and then applying a mask to the fully sampled $k$-space data with a sample rate of 1/4. We register the head CT and MRI images and use 17 paired multi-modality image pairs as training data. SimpleITK Python package~\cite{yaniv2018simpleitk} is used for image registration. Specifically, we optimize affine registration transforms via sampling with a rate of 0.01 and gradient descent with learning rate as 1.0 and number of iterations as 100. The CT modality is set as the fixed modality and the MRI modality is set as the moving one. For CT images, we perform window transform and normalize data to 0-1. We use 15 paired volumes for training and two paired volumes for inference. Based on the volumes, in total, we acquire 484 image pairs among which 95 pairs are reserved for inference. The OOD test dataset GLIS-RT is available through the Cancer Imaging Archive (TCIA)\footnote{\url{https://wiki.cancerimagingarchive.net/}}.

The MSL model was trained with 100,000 iterations. In each iteration, the model was fed with a batch of data (batch size set to 4). The learning rate started from 2e-4 and decayed by 0.5 every 10000 iterations. We used the Adam optimizer with the betas (0.5, 0.99) and weight decay 1e-4. Additionally, the hyperparameters of the loss objective were set as follows: $\phi_f$ = 1.0, $\phi_a$ = 1.0, and $\phi_e$ = 0.5.

\subsection{Results of MSL Imaging}
The input to our MSL model shown in Fig.\ref{overview}C is under-sampled CT and MRI sensory data, and the output is a sequence of MSL images, in which the global or local level of hybridization between CT and MRI is adjustable. By transferring crossover inter-modality information between CT and MRI, the image reconstruction of one modality can be enhanced or even completely done by the sensory data of the other modality.

\subsubsection{Qualitative and quantitative analyses of the MSL images}
The MSL model is trained with CT and MRI sensory data generated from 389 pairs of CT and MRI slices. For testing, we collected 95 pairs of CT and MRI slices from unseen subjects. 
Fig.~\ref{results}A shows the MSL images at three different slices for two test subjects. The images generated from the sparse sensory data with different levels of hybridization of CT and MRI, including the single-modality images ($\lambda$ = 0 and 1.0 for MSL-CT and MSL-MRI, respectively), are displayed along with the ground truth. At the top-left corner of each sample, the values of quantitative metrics including Mean Absolute Error (MAE), Structural Similarity Index (SSIM), and Mutual Information (MI) for the MSL images against the corresponding ground truth images are shown. SSIM is widely used to evaluate the structural similarity between images~\cite{b22}. MI measures the amount of information contained in one image about the other. It is observed that the sparse images (i.e., the images obtained by direct filtered back projection (FBP) reconstruction and inverse Fourier transform with sparse data) exhibit strong artifacts due to the severe under-sampling of the sensory data. On the other hand, the CT and MRI images generated by the MSL model are of high quality with high similarity to the ground truth. In Fig.~\ref{results}A, we also display the difference between the MSL image and the ground truth CT and MRI. It is seen that the MSL images with different $\lambda$ afford variable hybridization of CT and MRI and allow us to take advantage of the useful properties of the two modalities. For instance, more detailed information on the cerebral ventricles and tumors can be obtained with a larger $\lambda$, whereas the visualization of the skull and other body structures can benefit from a smaller $\lambda$. It is remarkable that, in the first case, the orbital features can be seen clearly in the MSL images because of the seamless integration of the bony structures of the CT and the soft tissue details of the MRI. Similar success is also seen in the second sample with complex structures and ample anatomical details. In the situation with a tumor (the third example), it is observed that the MSL images clearly reveal the tumor volume in the context of other anatomical structures in the brain. An online video showing the change in image content with $\lambda$ can be found on our project page \url{http://msl-demo.github.io/}.

Fig.~\ref{results}B summarizes the values of the evaluation metric (MAE, SSIM, and MI) averaged over the MSL images of 95 test slices. The metric values, which measure the difference of MSL image of different $\lambda$ from the ground truth CT and MRI images, are shown in the left and middle panels, respectively. It is seen that as $\lambda$ increases, the MAE towards the CT ground truth decreases, while the SSIM and MI increase. Meanwhile, the metric towards the MRI ground truth shows the opposite behavior.

\subsubsection{MSL imaging with locally tunable contrast}
A useful feature of our MSL framework is allowing for the reconstruction of images with a pre-defined spatial distribution of the hybridization parameter $\lambda$. Fig.~\ref{lambda}A shows the MSL images generated with six different distribution maps, showcasing the ability to tune the contrast of imaging content to highlight either soft tissue or bony structures through a spatial distribution of $\lambda$. Moreover, the spatial distribution of $\lambda$ can be automatically optimized to maximize the informative content of the resultant MSL image (e.g., the mutual information and entropy of the resultant MSL image. The formula (12) can be seamlessly replaced with other specifically designed optimization objectives tailored to specific use cases. Fig.~\ref{lambda}B shows the optimized spatial distribution map of $\lambda$ and the corresponding MSL images for the three samples in Fig.~\ref{lambda}A. MSL images better highlight the imaging contrast, specifically the bony structures in the first two cases and the tumor regions in the third case, without the introduction of human priors. The online video (\url{http://msl-demo.github.io/}) shows more examples.

\begin{figure*}[t]
\centering
\includegraphics[width=0.8\textwidth, height=0.3\textwidth]{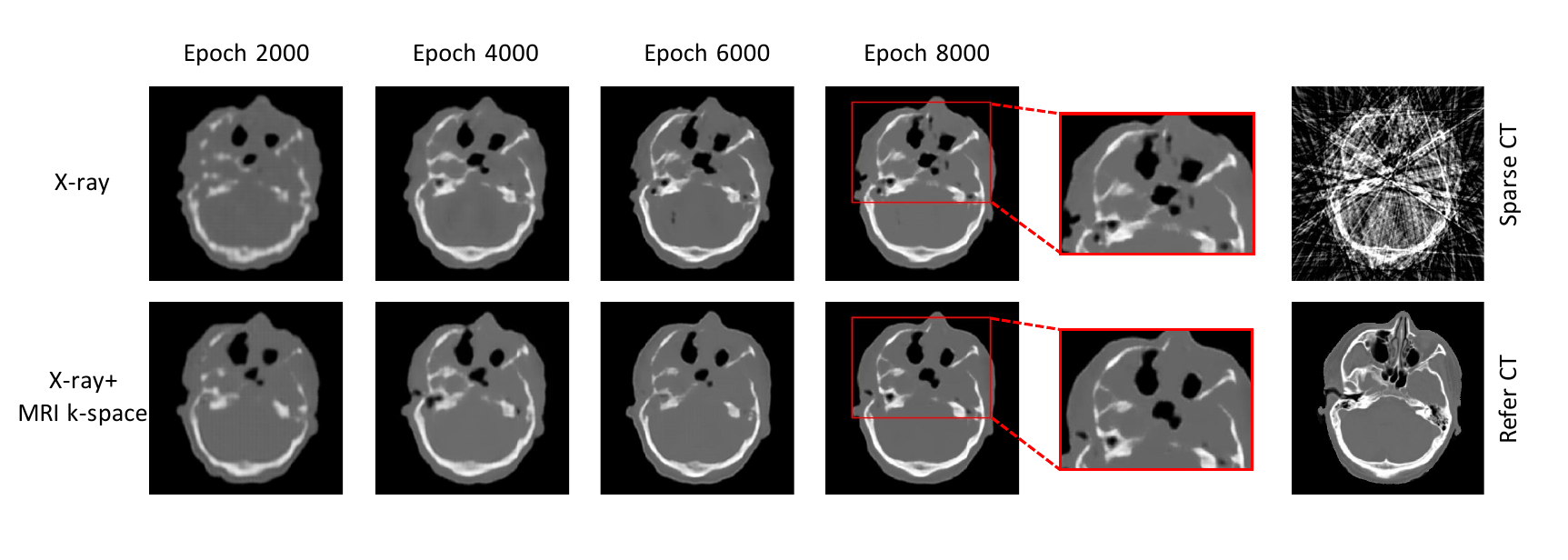}
\caption{\textbf{MSL enhances CT reconstruction using MRI k-space data.} 
The MSL-CT images obtained from sparse X-ray projections with (bottom row) and without (top row) the input of MRI k-space data are presented. Reconstruction results on different epochs are shown. The presence of MRI sensory data reduces the appearance of artifacts in bony structures in the MSL-CT images.} 
\label{crossover}
\end{figure*}

\begin{figure*}[t]
\centering
\includegraphics[width=1.0\textwidth, height=0.55\textwidth]{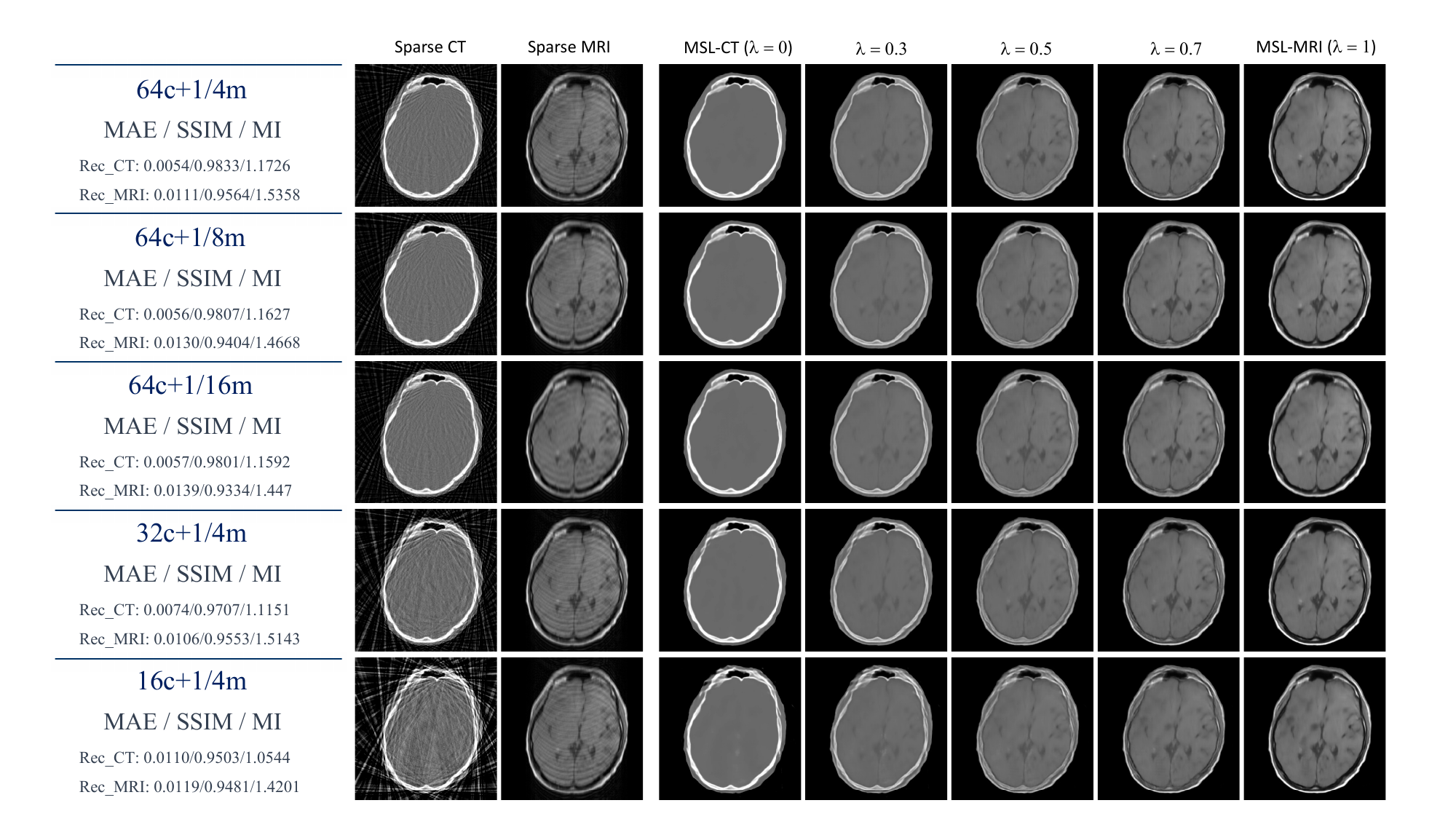}
\caption{\textbf{Robustness study of MSL imaging results against different sampling rates of sensory data.} 
Here “64c+1/4m” denotes 64 projection views for CT sinogram and an under-sampling rate of 1/4 for MRI $k$-space data. We show the sparse data input and the MSL images for each case.} 
\label{diffpara}
\end{figure*}

\begin{figure*}[t]
\centering
\includegraphics[width=0.65\textwidth, height=0.6\textwidth]{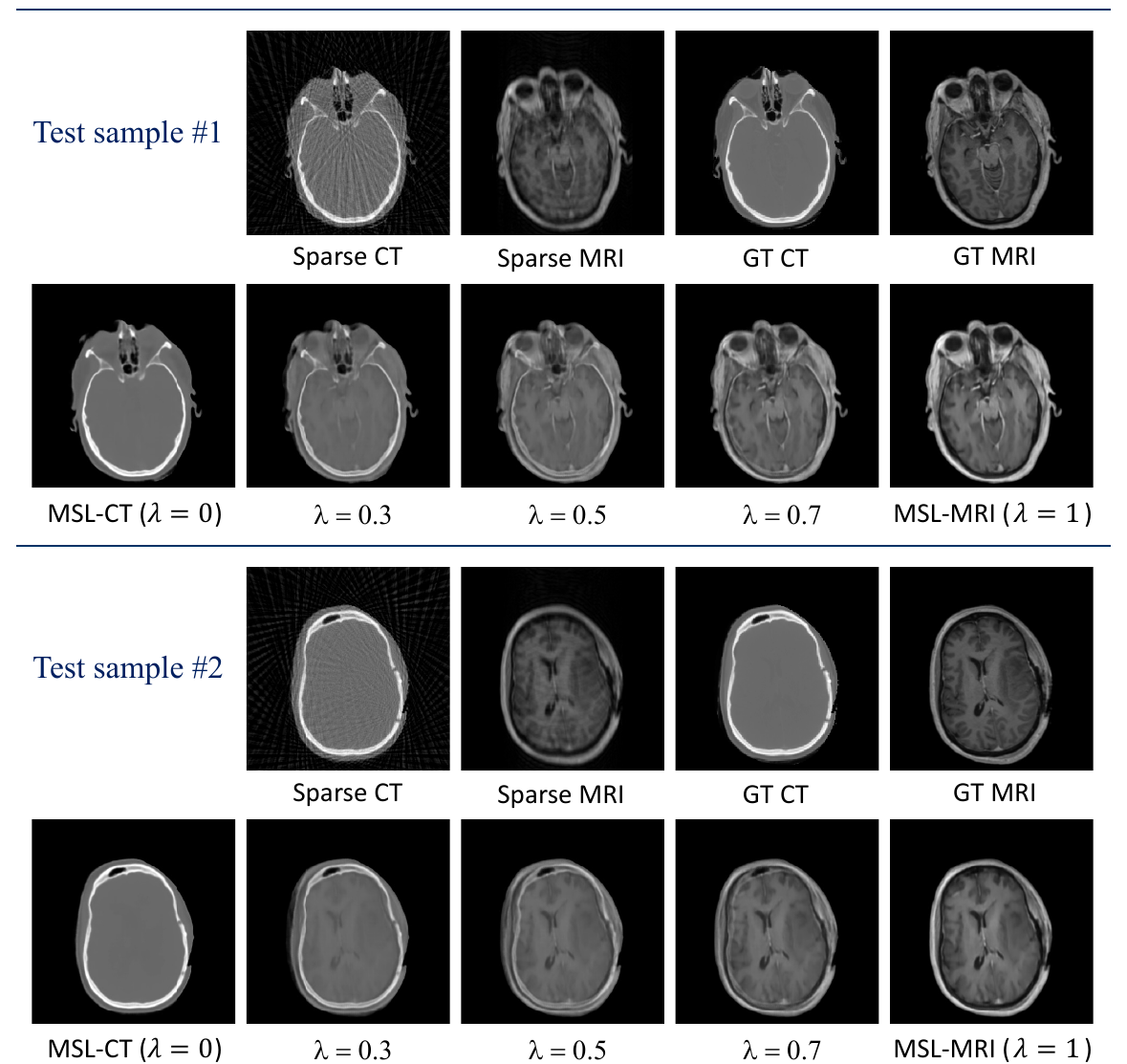}
\caption{\textbf{MSL imaging results on the out-of-domain test dataset.} 
Our model, trained on the RIRE dataset, is used to directly reconstruct MSL images from an unseen dataset.} 
\label{ood}
\end{figure*}

\subsubsection{MSL imaging from single-modality sensory data}
As illustrated in Fig.~\ref{overview}C, the transformer blocks produce intra-modality and inter-modality features from each input sensory data, termed $f_{intra\_CT}$ and $f_{inter\_CT2MRI}$ for CT, and $f_{inter\_MRI2CT}$ and $f_{intra\_MRI}$ for MRI. After cross-domain interaction, both CT and MRI representations gain extra information from the other modality, leading to enhanced CT and MRI images. It is worth highlighting that our MSL model can yield a continuum of hybrid CT/MRI images even when there is only one type of input sensory data. For instance, for input with MRI sensory data, only $f_{intra\_MRI}$ and $f_{inter\_MRI2CT}$ are generated. The MSL model uses $f_{intra\_MRI}$ features for the generation of MSL-MRI image, and the $f_{inter\_MRI2CT}$ features for generating MSL-CT. Fig.~\ref{onlymri} shows the MSL images generated by the model when the input is only MRI sensory data in the testing phase, and the resultant MSL images can preserve the soft tissue contrast while providing valuable bony structure information. 
In each of the above two special cases, the inter-modality features are the only source of information for the generation of MSL images beyond the input data domain. It is useful to note that, in the absence of CT or MRI sensory input data, the extra information needed to generate the hybrid images is from the prior knowledge casted in the trained MSL model. 

In the context of single-modality scenarios, we extend our comparison to include image translation methods. To benchmark our technique against existing image style transition methods in image domain, we chose the classical GAN-based Pix2pix model~\cite{p2p} as our baseline, which has been utilized for many image style translation tasks~\cite{richardson2021encoding, huang2018multimodal, li2021image}. Specifically, we take sparse MRI as input and use the Pix2pix to generate the corresponding CT images. The results from this model can be juxtaposed with those from MSL imaging under both multi-modality and single-modality conditions. It is worthy to note that, our method does not necessitate re-training for different input modalities; the same model can adapt to various input configurations. As seen in Fig.~\ref{translation}, though there is a moderate performance decrease when MSL is applied with only sparse MRI data compared to the multi-modality setting, our approach still outperforms Pix2pix, especially considering that the latter requires dedicated models for each specific task.

\begin{figure*}[t]
\centering
\includegraphics[width=0.65\textwidth, height=0.55\textwidth]{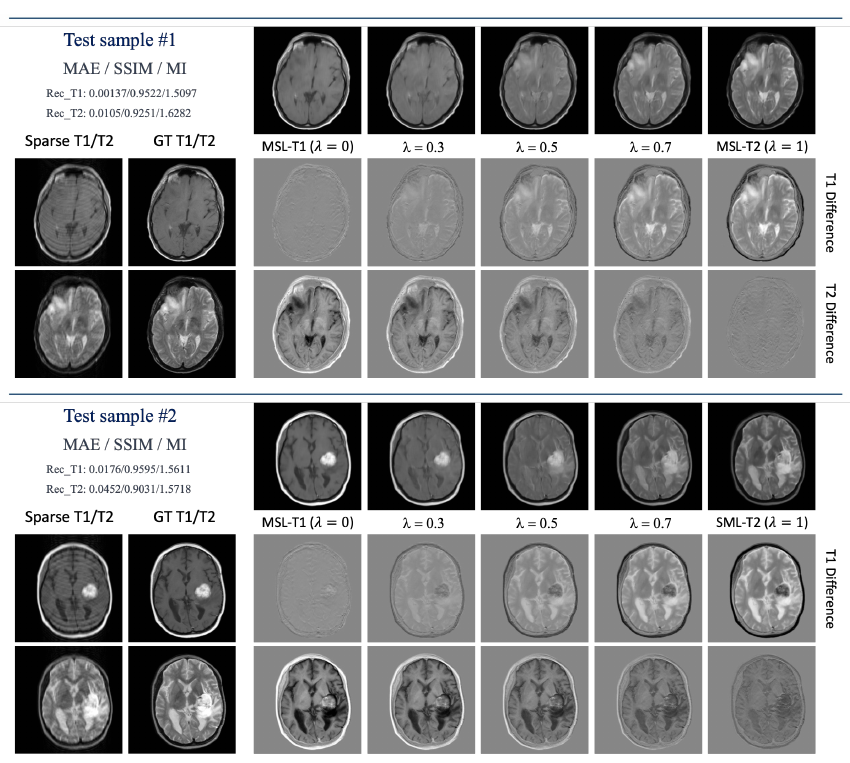}
\caption{\textbf{MSL imaging of $T_1$- and $T_2$-weighted MRI. } 
The reconstructed MSL images demonstrate the feasibility and effectiveness of the proposed MSL framework for multi-sequence MRI imaging.} 
\label{t1t2}
\end{figure*}

\subsection{Ablation Analysis of Our Framework}
\subsubsection{Crossover inter-modality information enhances MSL-CT and MSL-MRI image quality}
To further illustrate the impact of crossover inter-modality information transfer, we train two MSL models to generate MSL-CT images (i.e., $\lambda$=0) for sparse X-ray projections input (16 projection views) with and without under-sampled MRI $k$-space data. Fig.~\ref{crossover} shows the CT images generated by the two models at different epochs for a test case. For comparison, we also show the image reconstructed using the conventional FBP method and the ground truth CT image. With the cross-domain interaction module in our framework, the extracted crossover inter-modality information from MRI sensory data is transferred to enhance the CT reconstruction. From Fig.~\ref{crossover}, it is observed that MSL-CT with MRI sensory data consistently outperforms its counterpart without the MRI sensory data at different epochs. Additionally, we find that it becomes increasingly difficult to train a single-modality reconstruction model as the number of X-ray projections reduces, which is not the case for MSL-CT reconstruction in the presence of MRI sensory data. Furthermore, a comparison of the MSL-MRI reconstruction results (the rightmost column) in Fig.~\ref{results}A (with both CT and MRI sensory data input) and Fig.~\ref{onlymri} (with only MRI sensory data input) also demonstrate that the crossover inter-modality information of CT sensory data can improve MSL-MRI image quality. The above results provide pieces of evidence that our MSL framework can utilize the crossover inter-modality information for enhancing multi-modality imaging.

\subsubsection{Robustness study of MSL framework against different sampling rates of sensory data}
Data-driven image reconstruction is normally done with a model trained on a specific form of input data, which must be retrained when the input data format changes. In contrast, due to the transfer of information across modalities enabled by the inter-modality feature extraction, our MSL presents a generally applicable framework that can generate MSL images even when the sampling rate of CT and/or MRI sensory data is altered in the testing phase. For example, we train an MSL model with sparse CT (64 projection views) and MRI (an under-sampling rate of 1/4 of $k$-space data) sensory data. The trained model is used to generate MSL images with more sparse sensory data. Fig.~\ref{diffpara} shows the resultant MSL images with different sparsity of sensory data for one case. On the left, the sampling rate of the sensory data is specified.
By “64c+1/4m”, for example, we mean that the CT sinogram is from 64 views and the under-sampling rate of MRI $k$-space is 1/4. It is seen that our method can generate high-quality MSL images even with more sparse sensory data and the quantitative evaluation results on the left side of each row are close to that of the original sampling rate (i.e., 64c+1/4m). Unsurprisingly, a higher under-sampling rate of MRI and/or a smaller number of CT project views degrade the image quality. In the ultra-sparse limit, the artifacts appear in soft tissue and cerebral ventricles area, especially in the case “32c+1/4m” and “16c+1/4m”.

\begin{figure*}[h]
\centering
\includegraphics[width=0.65\textwidth, height=0.35\textwidth]{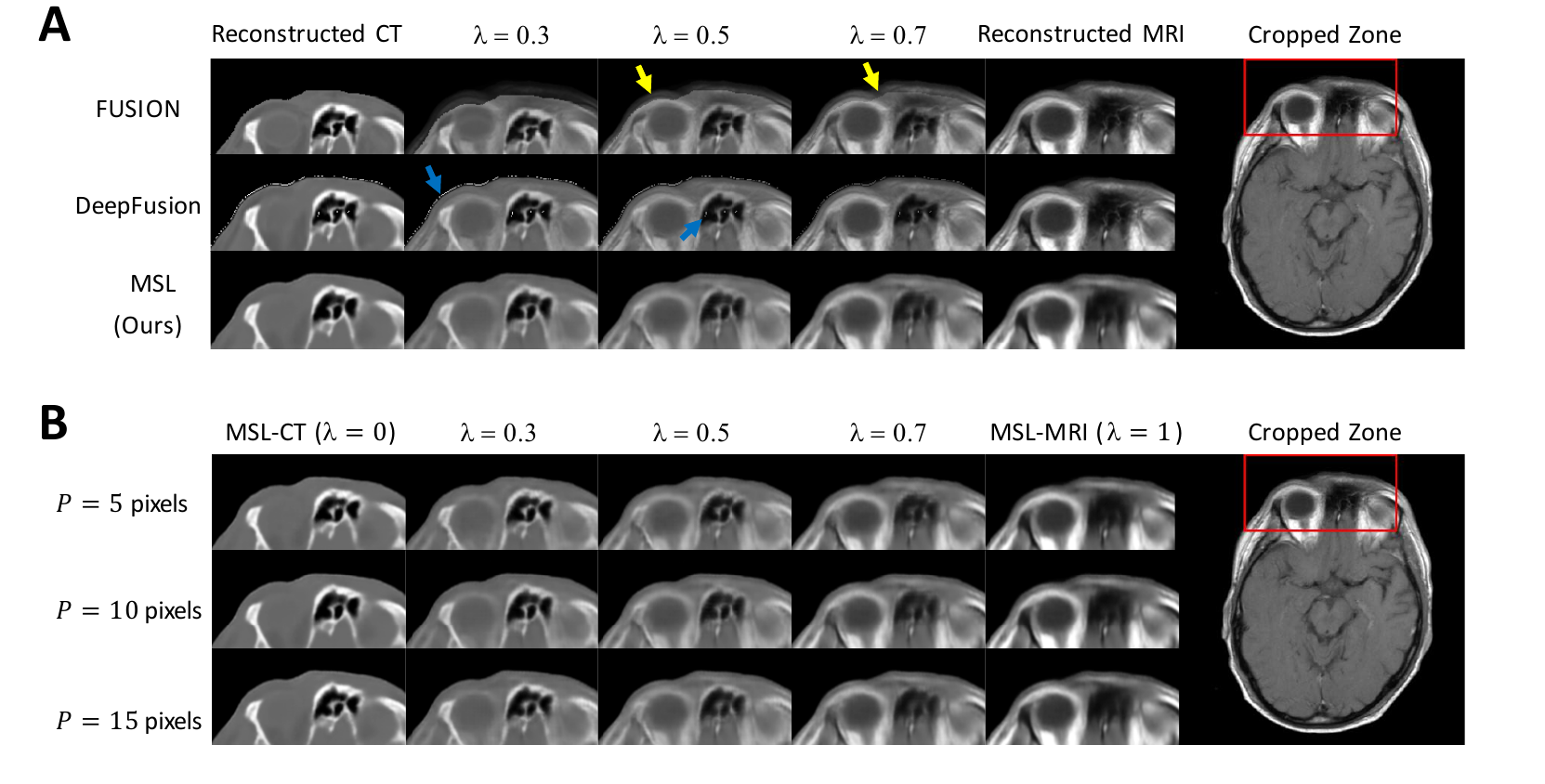}
\caption{\textbf{Robustness of MSL imaging against registration accuracy.} \textbf{(A)} Comparison of MSL imaging results with two baseline results. P=10 pixels for all three rows. \textbf{(B)} MSL imaging is less affected by registration inaccuracies of CT and MRI data, performing well under three different levels of perturbation. The red rectangle highlights the cropped zone near the eyes to better observe the differences.} 
\label{s3}
\end{figure*}

\subsubsection{Robustness of MSL imaging against registration inaccuracy of input CT and MRI data}
The resultant images from the proposed MSL framework are found to be less susceptible to registration inaccuracy of input CT and MRI data compared to post hoc fusion-based approaches. To illustrate this, we introduce small perturbations in the registered input CT and MRI data. Specifically, we randomly shift the CT images by $P \in (5,15)$ pixels in x- and y-directions and generate input sensory data. We then generate MSL images with the perturbed CT and MRI sensory data using our framework and compare the results with two baseline methods: (1) FUSION, which directly fuses the misaligned CT and MRI ground truth images in a pixel-wise fashion and takes different weights (i.e., corresponding to hybridization parameter $\lambda$) of CT intensity and MRI intensity to generate a series of hybrid images, and (2) DeepFusion (deep learning-based fusion), which utilizes two encoder-decoder-based models to reconstruct high-quality CT and MRI images from input CT and MRI sensory data, respectively. Meanwhile, we encourage the CT reconstruction model to wrap the CT image into the space of the MRI image and then fuse the reconstructed CT and MRI images in a pixel-wise fashion as the sequence of the hybrid images.  Fig.~\ref{s3}A shows the detailed MSL images generated by our framework (the third row) along with the two baseline results (the first and second rows). FUSION generates unsatisfied results when the data is not well registered, with artifacts in the edge area and shadow in the result (see yellow arrows). Although DeepFusion can capture the data perturbation via a deep reconstruction model, the resultant images have some noisy points (see blue arrows) and its performance is unsatisfactory. In contrast, our MSL approach can well capture the data perturbation and utilize cross-modality information to reconstruct the MSL from the perturbed data, resulting in improved image quality compared to the baseline methods. Moreover, we show the MSL images under different levels of perturbation in Fig. ~\ref{s3}B and the MSL performs well under all levels.

\subsection{Evaluation of Unseen Domain Data}
 Besides demonstrating robustness against different sampling rates, our MSL framework is also generally applicable to different domains and shows its generalization and utility. Specifically, our MSL framework can reconstruct images from out-of-domain (OOD) sources (e.g., data from another centre or equipment). This capability is crucial for data-driven imaging, as it allows high-quality imaging of data not seen during training. Fig.~\ref{ood} shows the generated MSL images on the GLIS-RT dataset~\cite{b24,b25} obtained from the Cancer Imaging Archive (TCIA)~\cite{b26} using the model trained with the RIRE dataset. Note that the GLIS-RT dataset is used as the OOD test dataset without re-training our MSL model. These results suggest that our MSL framework is applicable to different domains.

\subsection{The Synergetic MSL Imaging of $T_1$- and $T_2$-weighted MRI}
The proposed MSL strategy provides a generic method for synergetic multi-modality imaging, making it extendable to many other multi-modality imaging applications. In this subsection, we demonstrate the feasibility and effectiveness of using the MSL framework for synergetic imaging of $T_1$- and $T_2$-weighted MRI. To accomplish this, we train an MSL model using the $T_1$- and $T_2$-weighted MRI sensory data from the RIRE project as two input channels (see Fig.~\ref{overview}C). The resulting synergetic MSL images of $T_1$- and $T_2$-weighted MRI are shown in Fig.~\ref{t1t2}. It is evident that the reconstructed MSL-$T_1$ and MSL-$T_2$ images are of high quality, closely resembling the ground truth. Additionally, the reconstructed MSL images with different $\lambda$ provide variable hybridization of $T_1$ and $T_2$, enabling us to leverage the advantageous properties of both MRI sequences.

\begin{figure*}[t]
\centering
\includegraphics[width=0.8\textwidth, height=0.3\textwidth]{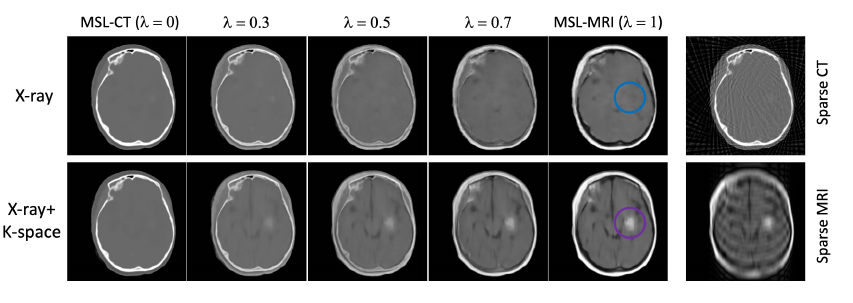}
\caption{\textbf{MSL imaging results with different information sources.} 
 The MSL imaging is from sparse X-ray projections only (top row) and both sparse X-ray projections and MRI k-space data (bottom row). Blue and purple circles indicate the tumor regions. It is difficult to observe tumor in the MSL images with only sparse X-ray projection as input, while the tumor shows up in the MSL images if some MRI sensory data are included in the input.} 
\label{tumor}
\end{figure*}

\section{Discussion}

\begin{figure*}[h]
\centering
\includegraphics[width=0.6\textwidth, height=0.75\textwidth]{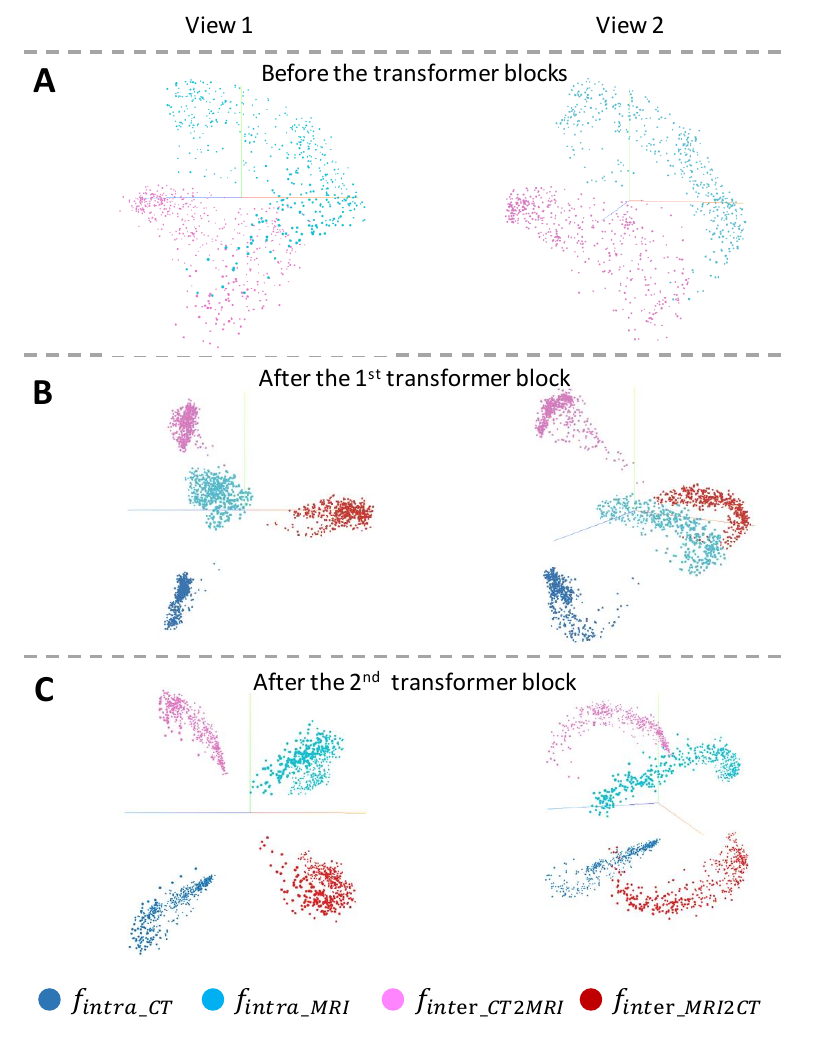}
\caption{\textbf{Low dimensional visualization of the extracted intra- and inter-modality features by MSL.} We reduce the features into 3D space using PCA and then project them onto a 2D space from different directions for visualization. Each color represents one type of features, and each point represents the feature of one sample. \textbf{(A)} Visualization before the transformer architecture. \textbf{(B)} Visualization at the middle layer of the transformer architecture (i.e., after the $1^{st}$ transformer block). \textbf{(C)} Visualization after the transformer architecture (i.e., after the $2^{nd}$ transformer block). } 
\label{s4}
\end{figure*}

\begin{figure*}[h]
\centering
\includegraphics[width=0.6\textwidth, height=0.55\textwidth]{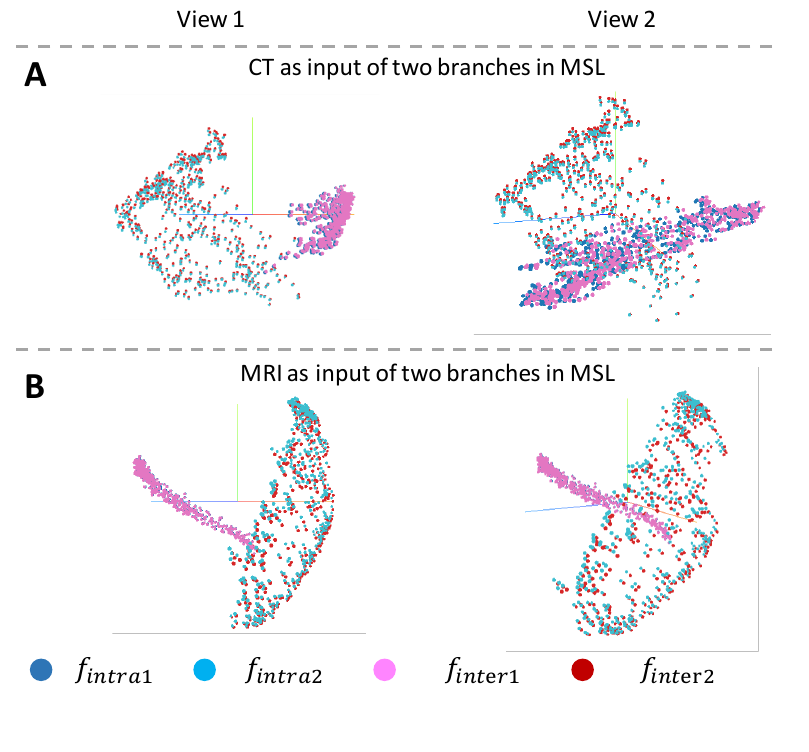}
\caption{\textbf{Low dimensional visualization of the extracted intra- and inter-modality features by MSL with the same CT/MRI sensory data as the input of both branches.} We display both intra-modality features ($f_{intra1}$ and $f_{intra2}$) and inter-modality features ($f_{inter1}$ and $f_{inter2}$) when feeding the same modality sensory data to the two input branches. \textbf{(A)} Visualization results with CT sensory data as input. \textbf{(B)} Visualization results with MRI sensory data as input.} 
\label{s5}
\end{figure*}

\begin{figure*}[h]
\centering
\includegraphics[width=0.65\textwidth, height=0.3\textwidth]{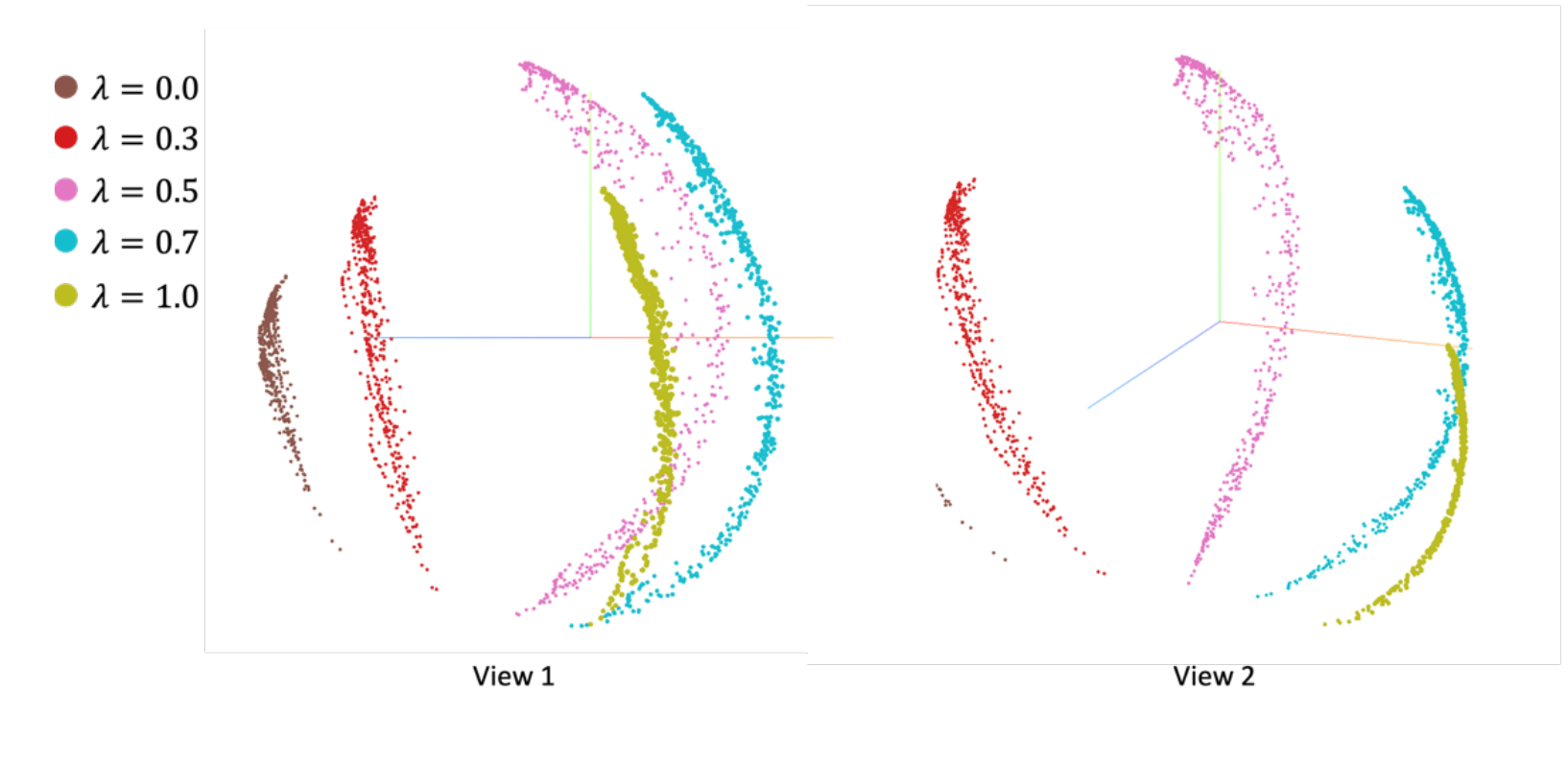}
\caption{\textbf{Low dimensional visualization of the features in the conditional decoder module of MSL.} Each color represents a feature with a specific hybridization parameter $\lambda$, and each point represents a feature of a sample.} 
\label{s6}
\end{figure*}

Data-driven multi-modality imaging (DMI) aims to enhance the contrast, resolution, and functionality of existing biomedical imaging techniques by combining different modalities and/or biomarkers for improved disease diagnosis and/or therapeutic guidance. Current DMI involves three main steps: data acquisition, image reconstructions of the involved imaging modalities, and post-reconstruction image fusion (Fig.~\ref{overview}A). While these three steps are inherently related, they are tacitly treated independently with the integration of multi-modality information done at the end by image fusion via either a software- or hardware-based registration method, leading to under-utilization of multi-sensor information. 
In this work, we have established a viable MSL framework to leverage the inter-relationship between different image modalities for augmented DMI. 
We found that the MSL-based image reconstruction of multi-modality sensory data enables information transfer across modalities and thus enhances the image quality of all involved modalities. The MSL breaks the boundaries of traditional imaging modalities and generates optimal hybrid images that would otherwise be impossible to attain.

In DMI, sensory datasets of different modalities are acquired to represent the same subject from different perspectives. It is, thus, intuitively understandable for different modalities to have some information overlap. It should be noted that, before the deep learning age, a joint reconstruction framework from CT and MRI data using compressed sensing techniques was studied by Yu and Wang~\emph{et al.}~\cite{b27}. In their approach, CT and MRI images were alternately reconstructed using the updated multimodal results that are calculated at the latest step of the iterative optimization algorithm. They noticed that by exploiting the structural similarities shared by the CT and MRI images, the similar image quality of fully sampled analytic reconstruction could be achieved with only a 20\% sampling rate for CT and 40\% for MRI. While the study was done before the data-driven age, it also emphasizes the role of cross-modality knowledge in mutually promoting the quality of image reconstruction. 
In our study, we also delve into learning the shared feature space that can be used to reconstruct both modalities. Within our MSL framework, we employ supervision from ground truth imaging data to guide the network to cultivate a compact latent representation space that can preserve the anatomical and pathological information of subjects. By utilizing the prior knowledge learned from the training data, the decoder in our framework can reconstruct high-fidelity CT/MRI or MSL images from this compact representation space. Moreover,  the assumption that high-fidelity image can be reconstructed from a latent code also aligns with the principles in previous auto-encoder based image reconstruction works. For instance, VQ-VAE~\cite{vqvae} is proficient in capturing intricate features within the latent space through an encoder-decoder framework, without the need for skip-connections. Similarly, \cite{miccai-vae} represents 3D MRI using 2D VAE models. When the latent space is well-trained to convey detailed information, it can be effectively utilized for reconstruction, even for multiple images as in our study. Our approach uniquely extracts the interplay between different modalities and capitalizes on the overlapping information to significantly augment the DMI process.

To gain a better understanding of the MSL framework and the role of transformer blocks in the cross-domain interaction module (see Fig.~\ref{overview}C), we visualize the intra- and inter-modality features extracted by the transformer blocks for brain imaging in Fig.~\ref{s4}. Specifically, for each training sample (slice), we compute the mean feature of each group of token features and use it as the inter-modality/intra-modality feature. Then, we use PCA to reduce the dimensionality of the four different types of features into three dimensions and project them into a 2D space for visualization from different directions. Each color corresponds to a type of feature, and each point represents the mean feature of a sample. We present visualizations of the features before the transformer blocks (Fig.~\ref{s4}A), in the middle of the transformer blocks (Fig.~\ref{s4}B), and after the transformer blocks (Fig.~\ref{s4}C) to highlight the evolution and relationship of these four different kinds of features.  
As we attribute the features of each modality into two groups before feeding them into the transformer blocks (e.g., duplicate $f_{CT}$  into $f_{intra\_CT}$ and $f_{inter\_CT2MRI}$), we can only observe two colors in Fig.~\ref{s4}A that correspond to the single sensory representations of CT and MRI. It is also observed that the features in the CT and MRI representations do not separate very well at this stage. However, at the middle layer of the transformer blocks (see Fig.~\ref{s4}B), the intra-modality features $f_{intra\_CT}$ and $f_{intra\_MRI}$ become more distinguishable, whereas the inter-modality feature $f_{inter\_CT2MRI}$ and $f_{inter\_MRI2CT}$ do not separate unambiguously. After the transformer blocks (see Fig.~\ref{s4}C), all four types of features are well separated into distinct clusters, indicating that our MSL framework is capable of learning different types of features (i.e., intra- and inter-modality features) from multi-modality sensory data. It is intriguing that the inter-modality feature $f_{inter\_CT2MRI}$ and $f_{inter\_MRI2CT}$ are located in between two intra-modality feature sets, suggesting their unique role in facilitating information transfer across different modalities.
To better understand the inter-modality features, we also trained an MSL model by feeding the same set of sensory data belonging to either CT or MRI to the two input branches of the MSL framework. As the two encoders are initialized with different weights, they extract different features even with the same input. Fig.~\ref{s5} shows the visualization of the extracted intra-modality and inter-modality features in this case. It is observed that there are now only two distinct groups of features corresponding to the information extracted by the two encoders. Not surprisingly, the inter-modality features are almost identical to the intra-modality features because of the use of only a single modality data for training the model. In this case, the model is not incentivized to learn the crossover information between them. We present the visualization of the features corresponding to different values of hybridization parameter $\lambda$ at the conditional decoder module of MSL in Fig.~\ref{s6}. For each training sample, we calculate the features for each different $\lambda$ and these features are displayed together in Fig.~\ref{s6}. Despite originating from the same multi-sensor representation, the features with different $\lambda$ are clearly distinguishable due to the CIN operations employed in the conditional decoder module. Thus, a sequence of MSL images is generated by varying the hybridization ratio.

MSL images are generated according to three sources of information: the input CT and MRI sensory data and the prior knowledge learned from the training data. In testing, depending on the available type of input sensory data, including (i) both CT and MRI, (ii) MRI alone, and (iii) CT alone, three different scenarios may occur and they all benefit from the proposed MSL strategy. In the first case, the inter-modality features from one modality would boost the other, leading to improved both CT and MRI images. In the second scenario, our MSL framework yields not only MRI but also CT images (Fig.~\ref{onlymri}). We emphasize that, in this case, the information needed for CT imaging arises from not only the inter-modality features extracted from the input MRI sensory data but also the training CT and MRI datasets. 
We note that the sensory data of different modalities are registered before feeding into the MSL model. Interestingly, we found that our framework is less sensitive to the registration inaccuracy of CT and MRI data, shown in Fig.~\ref{s3}, which relaxes the strict requirements for sensory data registration. 

As is well known~\cite{b28,b29,b30,b31,b32,b33,b34,b35}, the distribution of training data can have an impact on deep learning models. Specific to the third (similarly, the second) scenario above, for instance, the MRI information comes partially from the training datasets, in addition to the inter-modality features extracted from the input CT sensory data. When testing data contains information outside the training data domain, MSL may not be able to yield reliable image content. For example, if a tumor volume is not present in the input CT, it is unlikely to show up in the MSL images if the case is outside of the training data domain (i.e., if the training CT-MRI datasets do not contain any tumor volumes). However, interestingly, tumors can still appear in MSL images even if some sparse MRI sensory data of the patient are included in the input (See Fig.~\ref{tumor}). In practice, it is important to ensure the robustness of MSL imaging against this type of out-of-distribution problem for future clinical implementation. Previous research~\cite{b28} has suggested two potential solutions, including (i) the use of external models as a network add-on to detect out-of-distribution data and (ii) modification of the training-data distribution or the training strategy (e.g., by adding regularization, data augmentation, or leveraging adversarial training). Among various methods, the modification of the training-data distribution is arguably the most straightforward way to proceed. 

We would like to emphasize that the implication of MSL integration of multi-modality learning goes beyond simply synergizing two or more imaging modalities during image reconstruction. The strategy may also simplify the design of multi-modality hardware because of the reduced burdens of sensory data acquisitions, and enable optimal multi-modality imaging system design in the future. Specific to CT and MRI, the proposed DMI could allow for a novel hybrid design of CT and MRI scanners, maximizing the potential of both modalities. In reality, a fused hardware, capable of acquiring spatially registered MRI and CT sensory data, would be ideal for truly end-to-end optimized multi-modality imaging systems. We would like to point out that the feasibility of hybrid X-ray fluoroscopy and MRI imaging in a single exam without patient repositioning has been demonstrated by Fahrig~\emph{et al.}~\cite{b36}. Additionally, the rationale, feasibility, and realization of simultaneous CT-MRI have been highlighted in a perspective paper~\cite{b37}. Although this study only demonstrates the software feasibility of data-driven multi-modality imaging, with the recent success and extensive experience in the development of MRI-LINAC (linear accelerator), PET-CT, and PET-MRI systems~\cite{b38,b39}, it is likely that hardware-integrated CT and MRI coupled with data-driven MSL will emerge for truly optimal multi-modality imaging. This principle also inspires cross-modality medical images translation~\cite{zhu2023make, zhu2024generative}.
Finally, the proposed MSL strategy for exploiting the relationship of multiple imaging modalities is quite general and can be extended beyond the presented realm of CT-MRI imaging. In general, for multi-modality imaging with n modalities, we can use n different PI-encoders to extract the features from each sensory dataset and employ a similar transformer network (Fig.~\ref{overview}B \& C) to extract intra- and inter-modality features for all modalities. The framework then yields enhanced sensor representations by aggregating respective intra-modality and n-1 inter-modality feature sets, which are combined at the final stage to yield the hybrid MSL images. It is even possible to generate images of a modality in the absence of sensory data of that modality~\cite{b40}, similar to the case of hybrid CT-MRI imaging described in the previous section.

Several studies~\cite{nir-1,nir-2,nir-3, zhu2024deformable, zhao2024hfgs} have employed implicit neural representation learning or Gaussian Splatting~\cite{kerbl20233d} for reconstructing medical images. However, the adaptation of 3D representation for multi-modality applications remains an under-explored area. We posit that efforts to extend these representations to multi-modality scenarios, and their integration within our framework, hold significant potential. The fundamental principle of modeling the latent space through neural implicit representation could be shared, as the underlying physical projections for multiple modalities are inherently the same. Besides, another limitation of our existing attempts is the use of 2D images as input, with volume processing done slice by slice. Future works can be lifting our backbone to 3D within the similar framework. The challenge in extending the framework from 2D to 3D is mainly computational. Considering the use of 3D representations for the shared latent space may be a compelling direction. Treating the 3D space as a reflection of the physical world allows for the regularization of information from different modalities within this space, potentially leading to more accurate and robust image reconstruction and fusion. This could uncover novel solutions for reconstruction and fusion from multi-sensory data.

\section{Conclusion}
In this work, we proposed a novel deep learning-based multi-sensor learning (MSL) framework for augmented multi-modality imaging (DMI). Our MSL approach emerges as a promising solution for reconstructing multi-modality images from sparse sensory data, offering valuable insights for improving the design of future medical imaging hardware. We found that multi-modality imaging can be augmented by MSL through effective extraction and sharing of inter-modality information, providing a wealth of complementary information for enhanced clinical decision-making.
A broadly applicable MSL framework is established to maximally exploit the CT and MRI sensory data for DMI. This technique optimizes the hybridization of both input modalities at the individual pixel level, allowing us to take advantage of the useful features of both. Our MSL brain imaging studies reveal that the technique not only enhances all the involved image modalities but also provides patient-specific MSL images with optimal image-wide contrastive information.
Furthermore, MSL can effectively reconstruct the missing modality in scenarios where only single-modality sensory data is available, providing an effective way of synthesizing images from sparse data of a different modality. As a result of the reduced burden of data acquisition, the MSL strategy may open new avenues to simplify the design of hardware-based multi-modality imaging systems and lead to end-to-end optimized imaging solutions.
Finally, the proposed MSL strategy is quite broad and extendable to the development of new classes of multi-modality imaging techniques and promotes applications in other fields involving the integration of multi-sensory information.

\section*{Acknowledgments}
This work was supported in part by NIH under Grant 5 R01 CA256890, and Grant 1R01 CA256890617 and 1R01CA227713, in part by the Research Grants Council of the Hong Kong Special Administrative Region, China under Grant 17308321 and Grant T45-401/22-N, in part by Hong Kong Innovation and Technology Fund under Grant ITS/273/22.

\bibliographystyle{IEEEtran}
\bibliography{jrnl.bib}

\vfill

\end{document}